\documentclass[prb,twocolumn,superscriptaddress,showpacs,amsmath,amssymb]{revtex4}
\usepackage{amsfonts}
\usepackage{bm}
\usepackage{verbatim}

\usepackage{graphicx}

\def\Z{\mathbb{Z}}
\def\C{\mathbb{C}}
\def\R{\mathbb{R}}
\def\T{\mathbb{T}}
\def\k{{\bf k}}
\def\n{{\bf n}}
\def\m{{\bf m}}
\def\GL{{\rm GL}}
\def\U{{\rm U}}
\def\SU{{\rm SU}}
\def\BZ{{\rm BZ}}
\def\P{\check{P}}
\def\cH{\mathcal{H}}
\def\cP{\mathcal{P}}
\def\cC{\mathcal{C}}
\def\cQ{\mathcal{Q}}
\begin{document}
\title{Topology of the planar phase of superfluid $^3$He and bulk-boundary correspondence for three dimensional topological superconductors}
\author{Yuriy~Makhlin}
\affiliation{Low Temperature Laboratory, Aalto University,  P.O. Box 15100, FI-00076 Aalto, Finland}
\affiliation{Landau Institute for Theoretical Physics, acad. Semyonov av., 1a, 142432, Chernogolovka, Russia}
\affiliation{Moscow Institute of Physics and Technology, 141700, Dolgoprudny, Russia}

\author{Mikhail~Silaev}
\affiliation{Low Temperature Laboratory, Aalto University,  P.O. Box 15100, FI-00076 Aalto, Finland}
\affiliation{Institute for Physics of Microstructures RAS, 603950 Nizhny Novgorod, Russia}

\author{G.E.~Volovik}
\affiliation{Low Temperature Laboratory, Aalto University,  P.O. Box 15100, FI-00076 Aalto, Finland}
\affiliation{Landau Institute for Theoretical Physics, acad. Semyonov av., 1a, 142432,
Chernogolovka, Russia}

\date{\today}

\begin{abstract}
We provide topological classification of possible phases with the
symmetry of the planar phase of superfluid $^3$He. Compared to the
B-phase [class DIII in classification of A. Altland and M. R.
Zirnbauer, Phys. Rev. B 55, 1142 (1997)], it has an additional
symmetry, which modifies the topology. We analyze the topology in
terms of explicit mappings from the momentum space and also
discuss explicitly topological invariants for the B-phase. We
further show, how the bulk-boundary correspondence for the 3D
B-phase can be inferred from that for the 2D planar phase. A
general condition is derived for the existence of topologically
stable zero modes at the surfaces of 3D superconductors with class
DIII symmetries.
\end{abstract}
\maketitle

\section{Introduction}

Recently, topological classification for generic symmetry classes
of topological insulators and superconductors was given by
Schnyder et al.~\cite{Schnyder1Class} and
Kitaev~\cite{KitaevClass}. Depending on the presence and
properties of generic symmetries, namely the time-reversal and
particle-hole symmetry, ten symmetry classes can be identified,
and for each of them depending on the space dimensionality,
topological classification gives an integer invariant ($\Z$), a
binary invariant ($\Z_2$), or no invariant (only trivial,
`non-topological' insulators in this class). One example of a
topological insulator is provided by the time reversal invariant
B-phase of superfluid helium-3, which belongs to the DIII symmetry
class; a Hamiltonian in this class respects time-reversal and
particle-hole symmetries~\cite{Schnyder1Class,KitaevClass,Atland}
(see Section~\ref{subsec:ClassGen}). In thin films of superfluid
$^3$He the time-reversal invariant planar phase~\cite{VW} can
become stable. In this phase, the superfluid gap, isotropic in the
B-phase, is anisotropic and vanishes for the direction, transverse
to the film. Nevertheless, in 2D this system is gapful
(`insulating'). While this phase has not been identified
experimentally yet, in recent
experiments~\cite{Saunders1,Saunders2,SaundersSci13,SaundersPRL13,Levitin2014}
strong suppression of the transverse gap has been observed.

The planar phase has an extra discrete symmetry, a combination of spin
and phase rotations, which may modify the topological classification,
adding extra topological invariants.
In this paper, we set out to provide the topological
classification of insulators with this additional symmetry. We are
partially motivated by the above-mentioned
experiments~\cite{Saunders1,Saunders2,SaundersSci13,SaundersPRL13,Levitin2014}.
However, it is also interesting to see how additional symmetries
modify the results for one of the ten classes. Although in
principle additional (to time and charge reversal) unitary
symmetries can be dealt with in the Altland-Zirnbauer (AZ)
approach (cf.~Ref.~\onlinecite{ZirnbauerCommMathPhys2005}), we
also want to understand the relevant topology and topological
invariants directly in the explicit language of the homotopy
theory, that is by analyzing the homotopy equivalence of relevant
mappings (cf.~Ref.~\onlinecite{MooreBalents}). This allows one to
identify a topological class from the analysis of the topology of
the band structure (in the case of no disorder) in the bulk. Such
direct view is also of interest for the basic ten classes. We
begin with similar analysis for the B-phase (that is for the DIII
symmetry class) and reproduce the known results. The topological
classification of topological insulators in class DIII in
$d=1,2,3$ dimensions is rederived, again in the explicit language
of homotopy topology. We then account for the additional symmetry
of the planar phase and modify the classification accordingly. The
classification for the planar-phase symmetry in 2D is discussed,
for which Volovik and Yakovenko give an integer ($\Z$) topological
invariant in Ref.~\onlinecite{VolovikYakovenko1989}.

An extra motivation to study this particular case of the planar
phase is that we use it to construct a dimensional reduction for
general class-DIII topological superconductors. We show that the
topological properties of a 3D system and an embedded (2+1)D
system, which exist in any time-reversal invariant cross section
of the momentum space, are connected. As an application of such a
reduction we derive a generalized index theorem for 3D topological
superconductors, which provides an example of the bulk-boundary
correspondence in odd spatial dimensions.

We consider here a particular additional symmetry, which is realized in the planar phase. This
is a combination of a $\pi$ spin rotation around some axis followed by
a $\pi/2$ phase rotation. This symmetry is satisfied in superfluid $^3$He,
in which the spin-orbit interaction is very weak, but not necessarily in other
materials.
Nevertheless, the method of dimensional reduction, discussed below, can be applied to other classes
of topological superconductors and insulators
with various additional exact or approximate symmetries,
such as the point symmetry groups in crystals, cf.~Refs.~\onlinecite{TeoHughes2013,Ueno2013},
and pseudo-spin rotations in graphene.

\section{Topological classification for planar-phase symmetry}

\subsection{Parametrization and symmetries of the Hamiltonian}
\label{subsec:ClassGen}

Our considerations of the planar-phase symmetry are based on those for the DIII symmetry class,
defined below, and
we begin with the latter case. We consider non-interacting translationally invariant systems.
This allows us to characterize the system by a single-particle Bloch Hamiltonian $H(\k)$. Here
$H(\k)$ is a mapping from the momentum space to the space of Hamiltonians with certain constraints,
depending on symmetry.
This is easily extended to interacting systems using the Green-function formalism,
the effective Hamiltonian being given by the Green function at zero frequency, see,
e.g., Ref.~\onlinecite{Volovik2010}.
With such constraints, topologically (homotopically) non-equivalent mappings are
possible, and the goal of topological classification is to provide a complete list of the
equivalence classes of such mappings (two mappings are considered equivalent --- homotopic ---
if they can be continuously transformed to each other). In order to deal with the topology of the
mappings, let us first discuss their properties in more detail.

In a completely translationally invariant $d$-dimensional system $\k$ runs over the
infinite momentum space. In the latter case for a spherically symmetric system ($^3$He-B) at
$k\to\infty$ the Hamiltonian has a fixed matrix form (up to an inessential positive constant;
cf.~Sec.~\ref{sec:bbint}), and this allows one to compactify the momentum space by identifying all
the points at $k\to\infty$ to a single point, which reduces the $d$-dimensional momentum space
$\R^d$ to a sphere $S^d$.
Similarly, in a system with discrete translational symmetry (a periodic crystal, a topological
band insulator), the (quasi-)momentum space is the Brillouin zone (BZ), which is a $d$-dimensional
torus, $\T^d$. However, if we disregard the so called weak topological invariants and focus only on
the strong topological invariants robust to disorder~\cite{FuKane06,KitaevClass}, one can again
replace $\T^d$ by a sphere $S^d$ by gluing together all point at the BZ boundary. Thus, below we
consider mappings from the $d$-dimensional `spherical Brillouin zone'
$S^d_\BZ$. Below, we sometimes use the `spherical' language, assuming that $\k=0$ is
the north pole, and $k=\infty$ (or the boundary of BZ) corresponds to the south
pole. We further assume that opposite points, $\pm k$, correspond to the points on the
opposite ends of a same-latitude line (that is differ by a $\pi$-rotation about
the $z$-axis).

Further, the properties of the mapping are fixed by the conditions
that the Hamiltonian is hermitian, non-degenerate (i.e. has no
zero energy eigenvalues, a gapful spectrum) and by the symmetry
constraints. In class DIII of the ten-fold Altland-Zirnbauer
classification\cite{Atland} (see Refs.~\onlinecite{Schnyder1Class}
for a summary and further references), the system
--- the Bogolyubov-de-Gennes (BdG) Hamiltonian in the case of our
current interest --- possesses two symmetries,
which are the time reversal and charge conjugation. We consider a $4=2\times
2$-dimensional space of states (we show below that increasing this dimensionality does not modify
the result). This corresponds to the spin and two Bogolyubov
indices, each taking one of two values; below we use the Pauli matrices $\sigma_i$ and $\tau_i$ for
these indices, respectively.
In our notations below the BdG
Hamiltonian has the form
$\left(\begin{array}{cc}\varepsilon_\k &
\hat\Delta_\k\\\hat\Delta^\dagger_\k& -\varepsilon^{\rm T}_{-\k}\end{array}\right)$,
where
$\hat\Delta$ is a spin-symmetric and momentum-odd matrix (triplet pairing) of
the superfluid order parameter, while $\varepsilon_\k$ is the spectrum of
excitations in the normal state, e.g., $\varepsilon_\k=(\k^2/2m)-\mu$.
The simplest form of $\hat\Delta$ for $^3$He-B and the planar phase
is $\hat\Delta_B=(\Delta_B/k_F)(\sigma_x p_x + \sigma_y p_y+ \sigma_z
p_z)g_\sigma$ and $\hat\Delta_P= (\Delta_P/k_F)(\sigma_x p_x + \sigma_y p_y)g_\sigma$,
respectively.
Here $g_\sigma=i\sigma_y$.
The time-reversal symmetry (TRS) implies that
\begin{equation}
H(-\k)= g_\sigma H^T(\k) g_\sigma^{-1} \,. \qquad {\rm (TRS)}
\end{equation}
(We follow the notations of Schnyder et al.~\cite{Schnyder1Class}) The charge-conjugation symmetry
(or
particle-hole symmetry, PHS) also relates $H(\k)$ and $H(-\k)$, specifically,
$H(-\k)=-\tau_1 H^T(\k) \tau_1$.
Since TRS and PHS are antiunitary symmetries, for convenience we use their
combination, which basically constrains the structure of $H(\k)$ at each $\k$ (a chiral
symmetry):
\begin{equation}\label{cond:TRS-PHS}
H(\k) = - P H(\k) P \,, \quad P=\tau_1\sigma_y\,, \qquad {\rm
(PHS*TRS)}
\end{equation}
that is $H(\k)$ anticommutes with $P$. This implies that $H(\k)$ is block-off-diagonal
in the eigenbasis of $P$, and thus it is completely
defined by its block $M$ above the diagonal (the block below the diagonal being $M^\dagger$,
cf.~Ref.~\onlinecite{Schnyder1Class}). Equivalently and more specifically, one easily finds from
Eq.~(\ref{cond:TRS-PHS}) that $H$ is a real linear combination of 8 (instead of the initial 16)
terms:
$H = a\sigma_x+b\sigma_z+c\tau_1\sigma_x+d\tau_1\sigma_z+e\tau_2+f\tau_3
+g\tau_2\sigma_y+h\tau_3\sigma_y$.
Then, the TRS relates the values of these coefficients at $\k$ and $-\k$: the coefficients $f$, $g$ in
front of $\tau_3$ and $\tau_2\sigma_y$ are the same, while the other six change sign.
It is convenient to combine these eight real numbers, $a$, $b$, $c$, $d$, $e$, $f$, $g$, $h$ into
the following matrix:
\begin{equation}
M=\left(\begin{array}{cc}
A+iB & C+ iD\\C-iD & -A+iB
\end{array}
\right) \,,
\end{equation}
where $A=a-id$, $B=b+ic$, $C=h-ie$, $D=-g-if$.
Then the condition that $H$ is gapped
($\det H\ne0$ or, equivalently, $A^2+B^2+C^2+D^2\ne0$) reads
\begin{equation}
\det M \ne 0\,, \label{eq:det M}
\end{equation}
and the PHS maps $M\to-M^T$.

Thus, our problem reduces to finding the classes of topologically equivalent mappings $M:\BZ\to
\GL(2,\C)$ with the property
\begin{equation}
M_{-\k}=-M_\k^T\,. \label{eq:M-TRS}
\end{equation}
We refer to this property by saying that the mapping
$M$ is {\it odd}.

It is useful to further simplify the problem by reducing (continuously
tightening) $\GL(2,\C)$ to $\U(2)$ in a standard manner.
Namely, each $M\in \GL(2,\C)$ can be
uniquely presented as a product $M=\P U$ of a unitary $U$ and a positive hermitian $\P$ (polar
decomposition). Then $\P$ can be (e.g., linearly) retracted to the identity $\hat 1$.
Formally speaking, this means that  $\U(2)$ is a
deformation retract~\cite{Switzer,FFG} of $\GL(2,\C)$.
Thus, we
have to classify mappings $U: \BZ\to \U(2)$ with $U_{-\k}=-U_\k^T$ (one can easily
check all the details of this reduction).

\subsection{$^3$He-B: a $\Z_2$ invariant in 2D}
\label{sec:Bphase}

Each Hamiltonian is described by a unitary $U$, and we have to
classify mappings $\BZ\to \U(2)$ with the proper symmetries. To
find the classification, we recall that $\U(2)=(S^1\times
S^3)/\Z_2$; more specifically, each unitary $U$ can be presented
as a product of a phase factor and a special unitary matrix,
$U=e^{i\phi}S$, $S\in \SU(2)$ (i.e., $\det S=1$). This
presentation is not unique, since one can change simultaneously
the sign of both terms in the product. However, if we choose some
presentation at one point, say, $\k=0$, we can follow how the
phase factor and the matrix $S$ vary continuously over the
BZ.\footnote{More formally, each mapping $U:\BZ\to \U(2)$ can be
 lifted to the covering $S^1\times \SU(2)$ of $\U(2)$ ({\it lifting
 theorem}). Of course, we have to choose the image of one point,
 say, $k=0$, between two possibilities, but after that the covering
 mapping is uniquely defined.}

If we parameterize $S$ with a 4D unit vector $\m$ as
$S=m_0\hat 1+i(m_x\sigma_x+m_y\sigma_y+m_z\sigma_z)$,
the symmetry properties of $U(\k)$ (the odd parity) translate in the following:
considering $\det U(\k)$, we find that $e^{i\varphi_\k}$ is either equal or opposite to
$e^{i\varphi_{-\k}}$. Since they are equal at $\k=0$, we find that $e^{i\varphi_\k}$ is even:
$\varphi_{-\k}=\varphi_\k$. This implies that $S_\k$ is odd, which means that
$m_0, m_x, m_z$ are odd and $m_y$ is even:
 \begin{eqnarray} \nonumber
 m_0(-\k)=-m_0(\k), m_x(-\k)=-m_x(\k), \\
 m_z(-\k)=-m_z(\k), m_y(-\k)=m_y(\k)\,. \label{Sodd}
 \end{eqnarray}

Hence we have to classify mappings $e^{i\phi}:\BZ\to S^1$ and $S:\BZ\to
\SU(2)$ with these symmetry properties (\ref{Sodd}).

The first mapping is obviously always topologically trivial because it is even.

The mapping $S:S^2_\BZ\to S^3_{\SU(2)}$, however, may be nontrivial: because of the
symmetry relation (\ref{Sodd}), the points $\k=0$ and $k=\infty$ (the poles of $S^2_\BZ$)
are mapped to $\pm i\sigma_y$. Thus there is a topological $\Z_2$-invariant that shows
{\it whether they are mapped to the same or to different points}.

Clearly, both values of this invariant can be realized, and this gives a complete
classification, that is any odd mappings $\BZ\to S^3$ with the same value of
this $\Z_2$-invariant are homotopic to each other within the class of odd
mappings.
To see this, let us first consider the 1D case of the mapping $S:S^1_\BZ\to S^3_{\SU(2)}$ from a 1D
sphere (a circle). In this case one can consider the mapping of one half of $S^1_\BZ$ between the
north and south poles: (i) in the even case (with the zero $\Z_2$-invariant) this can be contracted
to its initial point, and the other half gets contracted as well by the odd symmetry; hence the
mapping is topologically trivial; (ii) in the odd case ($\Z_2$-invariant 1) this half starts at one
pole of $S^3_{\SU(2)}$ and ends at the other, and it can be deformed to any standard path between
the poles (e.g., a meridian line), and the other half gets deformed by symmetry to the other half
of the meridian; hence all odd mappings can be deformed to each other continuously. A similar
consideration can be found in the next section in more detail.

Going to the 2D case, one can imagine cutting $S^2_\BZ$ by a meridian circle in two hemispheres,
thinking of this circle as the 1D BZ. By first deforming this 1D BZ as above, and then gluing to it
the two 2D halves of the
2-sphere, we arrive at the conclusion above. Notice, however, that by going to $d=3$ and using the
same
procedure, one finds out that the 3D hemispheres of $S^3_\BZ$ can be attached to the 2D frame in
many topologically different ways, described by the degree of the mapping. Thus we can see that
in this case the (integer) degree of the mapping $S:S^3_\BZ\to S^3_{\SU(2)}$ is the only topological
invariant. In summary, in agreement with Refs.~\onlinecite{Schnyder1Class,KitaevClass} we find for
class DIII a $\Z_2$-invariant in 1D and 2D and a $\Z$-invariant in 3D.

Let us also remark that if in addition we assert that far away from the Fermi surface (i.e., `far
above and deep inside the Fermi sea') the Hamiltonian is dominated by the normal-state part $\propto
\tau_3$, so that we fix also the images of $k=0,\infty$ in
$S_\varphi^1$ (to $\pm1$), then we have an additional $\Z$-invariant that tells us, how
many `half-times' the image of $S^d_{BZ}$ encircles $S_\varphi^1$. In other words, this invariant
is given by the winding number of $\det M$ along an arbitrary path from $k=0$ to $k=\infty$,
which is $\int \mathop{\rm tr} (P H^{-1} dH)/(4\pi)$.

\subsection{Planar phase} \label{sec:planar}

As we have discussed in the introduction, the planar phase of
superfluid $^3$He has an extra symmetry as compared to $^3$He-B.
While in the bulk helium-3 only A- and B-phases are known to be
stable (in zero magnetic field), the planar phase may be
stabilized in thin films. In recent
experiments~\cite{Saunders1,Saunders2,SaundersSci13,SaundersPRL13,Levitin2014}
indications of the strongly distorted B-phase were found, and the
planar phase may become observable too. As estimated in
Ref.~\onlinecite{VolovikPlanar}, the superfluid gap, which is
isotropic in the B-phase in the bulk, becomes anisotropic in the
films with the gap in the transverse direction suppressed by a
factor of about 0.4.

 \begin{figure}
 \centerline{\includegraphics[width=0.8\linewidth]{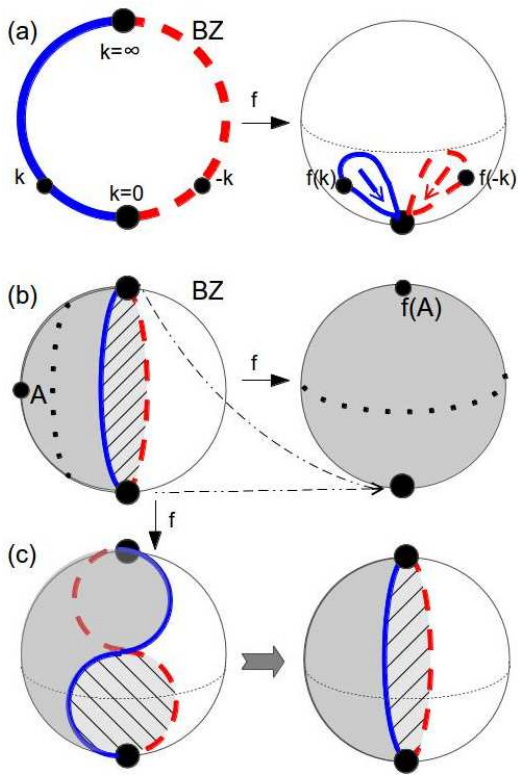}}
   \caption{\label{Fig:Maps} (Color online)
    The mappings $f$, analyzed for classification of the planar-phase symmetry class,
    are from the Brillouin zone (BZ), compactified to a sphere $S^d_\BZ$ with
the poles at $k=0$ and $k=\infty$, to the sphere $S^2_\n$ (see text).
    The mappings are odd: the images $f(\k)$ and $f(-\k)$ on $S^2_\n$ are related by a $\pi$
rotation around the (vertical) axis.
  (a) The 1D BZ --- a circle. For the zero value of the $\Z_2$ invariant both poles of BZ are
mapped to the same pole of $S^2_\n$. In this case $f$ can be continuously deformed to the trivial
mapping, which maps the whole circle to one point.
Indeed, the image of one semicircle (blue solid line) can
be continuously contracted to the pole as indicated by the arrow; the other semicircle (red dashed
line) is contracted at the same time, because the mapping is odd.
(b) The 2D BZ, a sphere, presented as a 1D frame (0 and $180^\circ$ meridians) and two
hemispheres. For the zero $\Z_2$ invariant, the 1D frame is contracted to a point
(dashed arrows). Then the image of one hemisphere covers $S^2_\n$ an integer number of times
(once in the figure), and for an odd mapping the other hemisphere --- the same number of times,
giving an even degree in total.
(c) For the non-zero $\Z_2$ invariant the two poles of BZ are mapped to two different
poles of $S^2_\n$. Then the image of one semicircle from the 1D frame (blue solid line) can be
deformed to a standard path (zero meridian). The image of each hemisphere in BZ covers the same
half-integer part of $S^2_\n$, giving an odd degree in total (degree one in the figure).
  }
 \end{figure}

This motivates us to analyze the topological classification for
symmetry classes with `extra' symmetries on top of the basic TRS
and/or PHS symmetries. On one hand, this can be analyzed within
the general frame of the AZ
approach~\cite{ZirnbauerCommMathPhys2005}. However, explicit
results for symmetries of interest and especially explicit
expressions for respective topological invariants are of great
interest (cf. the discussion for crystalline
solids~\cite{MooreAFM,Fu11,Zaanen,Furusaki13,HsiehFu}).

We analyze the symmetry class of the planar phase, that is the
DIII class with an extra symmetry described below, and provide a
complete classification. We show below that the $\Z_2$-invariant,
found in the previous section for a 2D class-DIII system, survives.
 Moreover, the complete classification within this symmetry class gives rise
to an integer topological invariant, with the $Z_2$ invariant
being its parity .\footnote{In other words, in going from the
planar phase to the general case, we lift the symmetry constraint,
and all even mappings become mutually equivalent, and so do the
odd mappings.} We explain these results below in this section.

The additional symmetry of the BdG Hamiltonian in the planar phase in our notation is $C=\sigma_z$
(that is, $C=\tau_0\sigma_z$),
\begin{equation} \label{Eq:PlanarPhaseSymmetry}
H(\k)=C H(\k) C\,,\quad C=\sigma_z\,,
\end{equation}
and we consider the symmetry class with this additional symmetry
constraint. Here $C$ is a combined $Z_2$ symmetry --- a
combination of the spin $\pi$-rotation about the $z$-axis and the
phase rotation by $\pi/2$. Note that the single-particle
Hamiltonian (and Green function) commute also with the
transformations generated by $C$, $\exp(i\alpha C)$, which form a
continuous ${\rm U(1)}$ symmetry group in similarity, e.g., to
spin rotations about $z$ generated by $S_z$. This, however, does
not impose additional constraints on $H$ and hence does not change
the topology of these single-particle quantities. Furthermore, the
many-body Hamiltonian and multi-particle quantities (e.g., the
two-particle Green function) obey only the discrete, but not the
continuous symmetry.

The Hamiltonian satisfying (\ref{Eq:PlanarPhaseSymmetry}) can
be transformed to the off-diagonal form with
\begin{equation}
M=-M^\dagger\,,  \label{eq:M-planar_sym}
\end{equation}
and thus
$U=-U^\dagger$. Hence $U$ is either $\pm i$ (the case of little
interest) or $U=i\n\bm{\sigma}$, a spin rotation by $\pi$ around
an arbitrary axis $\n$.
This second choice provides the nontrivial topology.
We have to classify mappings of
the sphere $S^2_\BZ$ to the sphere $S^2_\n$ that are odd:
opposite points are mapped to opposite points, and for both
spheres {\it opposite} refers to points, related by a
$\pi$-rotation around a specific axis (the $z$-axis for $S^2_\BZ$
and the $y$-axis\footnote{Of course, it does not matter for topological classification,
whether the axis is $y$ or $z$. Below we have in mind the $z$-axis
for both spheres, and refer to its ends as the north and south
poles.} for $S^2_\n$).

In complete analogy with the 3D case for class DIII above,
we find that in 2D such mappings are completely classified by the {\it degree of the mapping}, which
could assume any integer value. Moreover, whether this value is even or odd, is related to the
$\Z_2$ invariant, defined above. Indeed, both $k=0$ and $k=\infty$ in BZ are mapped to $\pm
i\sigma_y$, and the $\Z_2$ invariant above determines, whether they are mapped to the same or to the
opposite points of the $\n$-sphere. We show below that in the former case the degree is even and in
the latter case the degree is odd.

To prove this statement, in analogy to the previous section let us cut the BZ-sphere in two halves
with a line through $k=0$ and $k=\infty$, for instance, with $k_x=0$ (equivalently, in the language
of spherical BZ $S^2_\BZ$, with a full meridian circle on the BZ-sphere, for instance, the zero and
$180^\circ$-meridians, the blue solid and red dashed lines in Fig.~\ref{Fig:Maps}b). Because of
the odd parity the mappings of one hemisphere completely defines
the full mapping. However, the mapping of the hemisphere is constrained at the
boundary (the $0$ and $180^\circ$ meridians) --- the opposite
points of the boundary, $\pm \k$, should be mapped to the opposite points of $S^2_\n$. Such mappings
can always be thought of in the following way: (i) we have a mapping of one
half-meridian (say, the $0$ meridian) between the poles to $S^2_\n$ (with the ends mapped to the
same or opposite poles, depending on the $\Z_2$ invariant); (ii) the mapping of the other
half-meridian ($180^\circ$-meridian) is determined by the symmetry (odd parity); (iii) and the
mapping from the interior of the hemisphere somehow (arbitrarily) extends the mapping of the
boundary.

First, each mapping of the hemisphere can be continuously deformed under the constraint of odd
parity to a simpler mapping. Specifically, we can modify continuously the mapping of the
half-meridian (with the mapping fixed at its ends), the other half-meridian being modified in
accordance with the odd parity. One can
easily see that, because of the odd parity, this modification does not change the total area on
$S^2_\n$ covered by the image of the hemisphere in BZ (and again by odd parity, the other hemisphere
covers the same area). It is more convenient to describe this modification separately for two
possible values of the $\Z_2$ invariant.

In the case of the zero value of the $\Z_2$ invariant, when two ends of the zero meridian
are mapped to
the same pole (Fig.~\ref{Fig:Maps}a), the image of this meridian (a loop starting and ending at the
same pole) can be
continuously deformed to the pole itself. Then, the area spanned by the hemisphere is an integer,
and the total area covered by the mapping $S^2_\BZ\to S^2_\n$ is even. (Note that any integer
degree can be realized: to see that, one could just map the whole meridian to the
pole.) In the other case of odd $\Z_2$ invariant (Fig.~\ref{Fig:Maps}c), when two ends of the zero
meridian on $S^2_\BZ$
are mapped to the opposite poles, the image of this meridian can be continuously deformed to `just a
straight line', e.g., to the zero meridian on $S^2_\n$, as indicated in Fig.~\ref{Fig:Maps}c. Then,
the area spanned by the hemisphere is
half-integer, and thus the full sphere $S^2_\BZ$ covers $S^2_\n$ an odd integer number of times.

Thus {\it the degree of the mapping is the only invariant}. It can take any integer value. This
value
is even, when $k=0$ and $k=\infty$ are mapped to the same point, and odd, if they are mapped to
different (then opposite) points. These two cases correspond to two values of the $\Z_2$
invariant from the previous section.

A comment is in order on higher-dimensional matrices. So far we considered the
$4=2\times2$-dimensional Hilbert space of possible states. In general, in these symmetry classes we
can consider higher($2n\times2$)-dimensional spaces, with the 2D Bogolyubov-Nambu and
$2n$-dimensional `internal' space (spin and other degrees of freedom); large values of $n$ pertain
to realistic condensed-matter systems~\cite{KitaevClass}.
The symmetry conditions for
TRS, PHS, and the extra planar-phase symmetry then look the same as above
(cf.~Ref.~\onlinecite{Schnyder1Class}). The anticommutation with $P$ again implies the
block-off-diagonal form of the Hamiltonian, and Eqs.~(\ref{eq:det M}), (\ref{eq:M-TRS}),
(\ref{eq:M-planar_sym}) are valid. Hence, as above we have to classify odd mappings (\ref{eq:M-TRS})
$U\colon S^d_\BZ\to\U(2n)$. The analysis
follows the same route as above: for class DIII, first, presenting $U=e^{i\varphi}S$ with $\det
S=1$, we find that the map $e^{i\varphi}$ is always trivial; as for $S(\k)$, each of the poles
$\k=0,\infty$ is mapped to one of the two connected components of antisymmetric matrices from
$\SU(2n)$ with the pfaffian $\mathop{\rm Pf} S=\pm1$, and this defines a
$\Z_2$-invariant. As above from
direct homotopic-theoretical considerations, we see that for $d=1,2$ there are no other invariants,
while for $d=3$ the topological class is again fully characterized by an integer, the
$\Z_2$-invariant above being its parity. Thus we find the same results for topological
classification for class DIII ($\Z_2$ in $d=1,2$ and $\Z$ in $d=3$ dimensions).

Similar considerations apply for the planar-phase symmetry class,
in this case we again find a $\Z_2$ classification in 1D and $\Z$ in 2D as opposed to
a $\Z_2$ classification suggested earlier\cite{Roy}. Note that this
class contains $2n+1$ disconnected components: $iM$ ($iU$) is a
hermitian operator with $l$ positive and $2n-l$ negative
eigenvalues, where $l$ may vary from $0$ to $2n$ and is related to
the signature $(2l,4n-2l)$ of the hermitian operator $CH$. Each
component (the set of unitary anti-hermitian $2n\times
2n$-matrices $U$ with $l$ eigenvalues $-i$ and $2n-l$ times $i$,
i.e., the grassmanian $\U(2n)/\U(l)\times\U(2n-l)$), except
$l=0,2n$, has the same second homotopy group $\pi_2=\Z$, and hence
a $\Z$-invariant in 2D arises. The cases $l=0,2n$ correspond to
$U=\pm i$ as above for $n=1$ (see the discussion below
Eq.~(\ref{eq:M-planar_sym})).

\section{Index theorems in odd spatial dimensions}
\label{sec:bbint}

\subsection{The problem}
\label{subsec:problem}

We have found a complete set of topological invariants for the planar-phase symmetry and for the
B-phase symmetry, and further related questions need to be analyzed. In particular, it would be
useful to have an explicit (integral) expression for the invariants. A further question of great
current interest concerns the bulk-boundary correspondence between the topological invariants in
the bulk and the properties of gapless boundary modes.

Since the discovery of a topological invariant for the integer
quantum Hall state~\cite{Thouless} there has been great interest in
deriving index theorems that connect the topology of the
fully-gapped spectrum in the bulk with the number of
gapless modes at the boundary of the system or inside topological
defects (strings, domain walls, monopoles, etc.).

A full classification is still absent, though there is certain
progress in understanding for even spatial dimensions, especially
when  the bulk  system is characterized by an integer-valued
topological invariant of group $\Z$. In 2D one can mention three
representatives of such systems: the integer quantum Hall effect
state (class A according to the general ten-fold classification);
the $^3$He-A phase and chiral $k_x+ik_y$ superconductivity (class
D); and the planar phase of $^3$He.

Among systems in odd spatial dimensions, of particular interest is the topological superfluid
$^3$He-B, which belongs to class DIII according to the general classification
scheme. The $^3$He-B Hamiltonian and the related Hamiltonians have the following form:
 \begin{equation}
 \label{Eq:HeBHamiltonian-gen}
 \cH_{3D}=\tau_3\epsilon(\k)+
 \tau_1 \left[\sigma_x f_x(\k) + \sigma_y f_y(\k)+\sigma_z
 f_z(\k)\right] \,.
\end{equation}
This Hamiltonian is gapful, when ${\bf f}(\k)$ does not vanish at the Fermi surface, where
$\epsilon(\k)=0$.
For convenience, here and below we use the form of the BdG
Hamiltonian, corresponding to an alternative definition of the
Nambu spinor~ (see p.77 in Ref. \onlinecite{VolovikHeDroplet} for
definition); it differs from the standard form by a unitary
transformation,
\begin{equation}
 \label{Eq:transformation}
 \cH = U^\dagger HU ~~,~~ U=\frac{1+ \tau_3}{2} + i\sigma_y \frac{1- \tau_3}{2} \,.
\end{equation}
For clarity we use calligraphic letters for
operators in this form. For the $p$-wave $^3$He-B, the functions of
the 3-momentum $\k$ can be chosen as
 \begin{eqnarray}
 \label{Eq:HeBHamiltonian}
 &\epsilon(\k)=&{\bf k}^2/2m -k_F^2/2m, \\ \nonumber
 &{\bf f}(\k) =&f \k\,,
 \end{eqnarray}
  where $f>0$.
The Hamiltonian (\ref{Eq:HeBHamiltonian}) has a symmetry-protected
topological invariant (\ref{TopInvariant_B}) with $N_{\rm B}=2$.
The more general Hamiltonian (\ref{Eq:HeBHamiltonian-gen}) may
have any even invariant $N_{\rm B}$ under the conditions that
 $f_{i}(\k)/\epsilon(\k)\rightarrow 0$ at $k\rightarrow \infty$,
where $i=x,y,z$, and $\epsilon(|\k| \rightarrow \infty) >0$. These
conditions allow compactification of the momentum space to
$S^3$, but they are not needed in crystals, since in that case the Brillouin
zone is a compact space. An example of a nontrivial mapping with a
higher topological charge $N_{\rm B}=2n$ is
 \begin{eqnarray}\label{Eq:HigherInvariants}
 &f_z (\k)&=f k_z,\\\nonumber
 &f_x (\k)&  =f {\rm Re} (k_x \pm ik_y)^{|n|},\\ \nonumber
 &f_y (\k)&  =f {\rm Im} (k_x \pm ik_y)^{|n|},\\ \nonumber
 &\epsilon(\k)&=\mu [ ({\bf k}/k_F)^{2|n|}-1],
 \end{eqnarray}
 where $n\in \Z$ and the upper (lower) sign corresponds to $n>
(<)0$. The form of $\epsilon(\k)$ dispersion in
Eq.~(\ref{Eq:HigherInvariants}) is chosen in such a way to allow
compactification of momentum space.  Alternatively, higher
values of the topological invariant can be obtained in a system
consisting of several layers of the planar phase.

\begin{figure}
 \centerline{\includegraphics[width=1.0\linewidth]{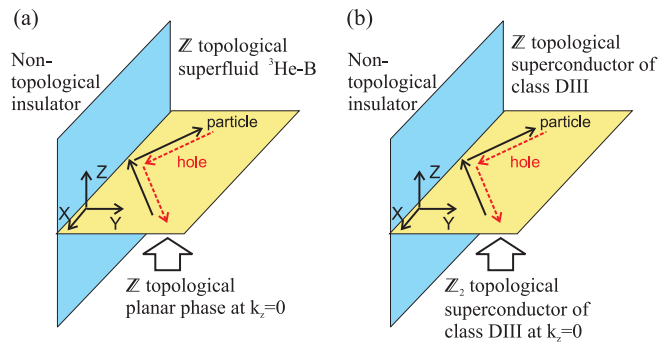}}
   \caption{\label{Fig:Reduction} (Color online)
  Dimensional reduction of the surface-states spectral problem in
  3D to that in a time-reversal invariant cross section of momentum space $k_z=0$.
  (a) Reduction from the
 $\Z$ topological superfluid $^3$He-B results in the $\Z$ topological planar phase at $k_z=0$.
  (b) Reduction from a general $\Z$ topological superconductor of class
  DIII produces a $\Z_2$ topological superconductor at $k_z=0$. }
 \end{figure}

In order to derive the index theorem for $^3$He-B
(\ref{Eq:HeBHamiltonian}) and related Hamiltonians
(\ref{Eq:HigherInvariants}) we assume that the boundary plane is
$y=0$, so that the conserved momentum projections are $k_{x,z}$.
To find the complete spectrum of bound states
$\varepsilon_b=\varepsilon_b(k_x,k_z)$ it is enough to consider a
set of 2D spectral problems for the cross sections of momentum
space
 \begin{equation}\label{Eq:Line}
  k_z \cos\theta + k_x \sin\theta=0 \,,
 \end{equation}
where $2\pi>\theta\geq 0$.
 Indeed the bound states at the interface between the
non-topological insulator and $^3$He-B are formed due to the
subsequent Andreev and normal reflections of particles and holes
as shown schematically in Fig.\ref{Fig:Reduction}. The momenta
of both the incident particle and the one reflected from the
boundary belong to the same cross section (\ref{Eq:Line}). Further,
we will use the fact that the planes determined by
Eq.(\ref{Eq:Line}) are time-reversal invariant in the sense that
they contain states with opposite momenta ${\bf k}$ and $-{\bf
k}$.

An example of such a dimensional reduction to the plane $k_z=0$ is
shown in Fig.\ref{Fig:Reduction}a.  The 2D Hamiltonian in this
cross section reduced from the 3D phase
(\ref{Eq:HigherInvariants}) is given by
 \begin{equation}
 \label{Eq:planarHamiltonian-gen}
 \cH_{2D}=\tau_3\epsilon(k_x,k_y) +  \tau_1 \left[\sigma_x f_x(k_x,k_y) + \sigma_y
 f_y(k_x,k_y)\right].
 \end{equation}

 The 2D Hamiltonian of the form
(\ref{Eq:planarHamiltonian-gen}) has the symmetry of the
 generalized planar state (\ref{Eq:PlanarPhaseSymmetry}).
In our current representation of the BdG Hamiltonians [see below
Eq.\ref{Eq:HeBHamiltonian-gen}], the operator of this additional
symmetry -- the matrix commuting with the
 Hamiltonian (\ref{Eq:planarHamiltonian-gen}) -- is  $\cC=U^\dagger C U=\tau_3\sigma_z$, with the
 unitary operator from Eq. (\ref{Eq:transformation}).
 The Hamiltonians satisfying this additional symmetry are classified by an
integer-valued topological invariant.
The explicit form Eq.(\ref{TopInvariant_P}) expressed via the Green function
 gives an even-valued topological invariant $N_{\rm P}$.
 For the particular set of parameters
(\ref{Eq:HigherInvariants}) this invariant is $N_{\rm P}=2n$. In
general it gives the Chern number $N_{\rm P}/2$ for each spin
projection and therefore yields an index theorem for the number of
edge states. Hence we conclude that $N_{\rm P}$ defines the
number of zero edge modes at $k_z=0$ in the parent 3D $^3$He-B
phase (\ref{Eq:HigherInvariants}) as well.

Below we show that the topological invariants for the Hamiltonians
(\ref{Eq:HeBHamiltonian-gen}) and (\ref{Eq:planarHamiltonian-gen})
coincide, $N_{\rm B}=N_{\rm P}$,  for quite a general form of the
order parameter with arbitrary functions $f_{x,y}({\bf k})$ and
$f_{z}({\bf k})\propto k_z$. Then the index theorem for $^3$He-B
states that the number of zero modes at the $k_z=0$ cross section
of the momentum space is given by the 3D invariant $N_{\rm B}$.
This particular choice of a cross section is determined by the
specific form of the $^3$He-B Hamiltonian
(\ref{Eq:HeBHamiltonian-gen},\ref{Eq:HigherInvariants}).
 Using this result we argue that for a general Hamiltonian of
class DIII the time-reversal invariant cross sections
(\ref{Eq:Line}) have a $\Z_2$ invariant, a non-zero value of which
protects at least one stable zero of the bound-state spectrum
$\varepsilon_b=\varepsilon_b(k_x,k_z)$ along  each line from
the one-parameter family (\ref{Eq:Line}).

We construct a map of the 3D
slice in the 4D momentum-frequency space to the space of
Green-function matrices $(k_x,k_y,t)\to \GL(4,\C)$
\begin{equation}\label{Eq:Map}
G=G(k_x,k_y,t,\alpha) \,,
\end{equation}
where $0<\alpha<2\pi$ is a parameter as shown in
Fig.\ref{Fig:Planes}. The map is designed to coincide with that
for the planar state at $\alpha=0$, when $t=\omega$, and for
$^3$He-B at $\alpha=\pi/2$, when $t=k_z$. They transform to each
other by a continuous change of the orientation of the 3D slice
in the 4D momentum-frequency $(\omega,{\bf k})$-space. We show
below that the homotopy class is independent of $\alpha$,
and this proves that the generalized 3D $^3$He-B is topologically
equivalent to the generalized 2+1 planar state in
the $k_z=0$ cross section.

\subsection{Dimensional reduction from $^3$He B to the planar state}

 \begin{figure}
 \centerline{\includegraphics[width=1.0\linewidth]{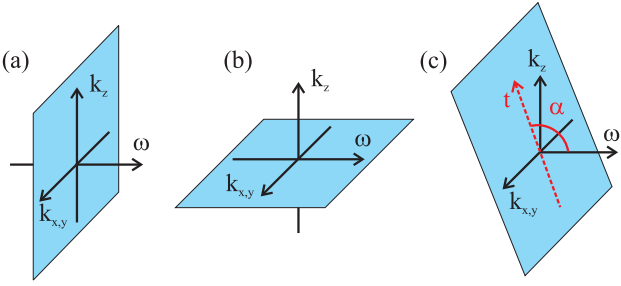}}
   \caption{\label{Fig:Planes} (Color online)
  Slices of the 3+1 dimensional space $(\omega, k_x,k_y,k_x)$ that are used to define the
  topological invariants for: (a) 3D $^3$He-B phase;
  (b) 2+1 dimensional planar phase; (c) smooth transformation between 3D
  and 2+1 dimensional invariants. The axillary axis is $t=\omega\cos\alpha+k_z\sin\alpha $.  }
 \end{figure}

We start with a particular case of the simplified Green function
describing the $(3+1)$-dimensional  topological superfluid:
 \begin{equation}\label{3He-B}
 G^{-1}(\omega,{\bf k})= i \omega - \cH_{3D} ({\bf k}),
 \end{equation}
 where $\cH_{3D} ({\bf k})$ is given by Eq.(\ref{Eq:HeBHamiltonian-gen}) with $f_z\propto k_z$.
There are two special cases: (i) $\omega=0$; (ii) $k_z=0$.

In the case  $\omega=0$, the Green function
$G^{-1}(0,k_x,k_y,k_z)$ represents a  $^3$He-B-like Hamiltonian,
which anticommutes with the matrix $\cP=\tau_2$, $\cP^2=1$
(chiral symmetry). The Hamiltonians with such symmetry  have the
following symmetry-protected topological invariant
\begin{equation}
N_{\rm B} = \frac{e_{ijl}}{24\pi^2} \mathop{\bf tr}
\int\limits_{\omega=0}   d^3k ~\cP ~G\partial_{k_i} G^{-1}
G\partial_{k_j} G^{-1} G\partial_{k_l} G^{-1}\,,
\label{TopInvariant_B}
\end{equation}
where $\cP=\tau_2$. Here the integral is taken over the momentum
space, i.e., over the $\omega=0$ slice in the 4D
$(\omega,k_x,k_y,k_z)$-space as shown in Fig.\ref{Fig:Planes}a.
The choice of slices in the frequency-momentum space of the
Standard Model is given in Ref.~\onlinecite{Volovik2010}. One can
check by a direct calculation that the invariant (\ref{TopInvariant_B})
is {\it twice} the $\Z$-invariant from Section~\ref{sec:Bphase} in
3D, that is the degree of the mapping from BZ to the 3-sphere of
$\SU(2)$ (it accumulates equal contributions from $M$ and
$M^\dagger$). For $^3$He-B parameters (\ref{Eq:HeBHamiltonian}) in
Eq.(\ref{3He-B}) one finds that $N_{\rm B}=2$ and all the higher
values of the invariant $N_{\rm B}=2n$, $n\in \Z$ are realized by the
set (\ref{Eq:HigherInvariants}).

In the case  $k_z=0$, Eq.(\ref{3He-B}) represents the  Green
function of the generalized planar phase
 \begin{equation}\label{planar}
   G^{-1}(\omega,k_x,k_y)=
 i \omega -\cH_{2D}(k_x,k_y)\,,
 \end{equation}
 where the Hamiltonian $\cH_{2D}=\cH_{2D}(k_x,k_y)$ is given by
  Eq.(\ref{Eq:planarHamiltonian-gen}).
The Green function of a 2+1 system with symmetry $\cC$ has a
symmetry-protected topological invariant,
which determines transport properties of the 2+1 system. Indeed,
it defines the quantized spin Hall conductivity in the absence of external magnetic
field~\cite{VolovikYakovenko1989}. The invariant  involves the symmetry operator
$\cC=\tau_3\sigma_z$, which commutes with the  Green function:
\begin{equation}
 N_{\rm P}= \frac{e_{ijl}}{24\pi^2} ~
\mathop{\bf tr} \int    d^2k d\omega ~\cC~ G\partial_{k_i} G^{-1}
G\partial_{k_j} G^{-1}G\partial_{k_l}  G^{-1}\,.
\label{TopInvariant_P}
\end{equation}
Here the integral is taken over the
$k_z=0$ slice, $k_i=(\omega,k_x,k_y)$, in the 4D
$(\omega,k_x,k_y,k_z)$-space as shown in Fig.\ref{Fig:Planes}b.
Again, one checks by a direct calculation that this invariant is
twice the $\Z$-invariant from Section~\ref{sec:planar} in 2D, that
is the degree of the mapping from BZ to the 2-sphere $S^2_\n$.

For particular cases, e.g., for the set of Hamiltonians with the
order parameter (\ref{Eq:HigherInvariants}), one finds that the
invariant (\ref{TopInvariant_P}) in the 2D cross section $k_z=0$
coincides with the value of the invariant in the parent 3D phase
(\ref{TopInvariant_B}), $N_{\rm P}=N_{\rm B}=2n$. This is not a
coincidence. Let us show that in a more general case  of
arbitrary functions $f_{x,y} ({\bf k})$ and $f_{z} ({\bf
k})\propto k_z$ the  even-valued integrals (\ref{TopInvariant_B})
and (\ref{TopInvariant_P}) can be continuously transformed to each
other by the rotation of the 3D slices in the 4D space, shown
schematically in Fig.\ref{Fig:Planes}.
 Such transformation allows us to connect topological properties
of the Hamiltonians in 3D and in 2D at $k_z=0$.

To construct the connection between topological invariants
(\ref{TopInvariant_B}) and (\ref{TopInvariant_P}) let us use the
Green function in the form $\tilde G^{-1}=-iG^{-1}\tau_2$:
\begin{eqnarray}
  \tilde G^{-1}(\omega,\k)=
  \omega \tau_2- \tau_3\sigma_z f k_z - \tilde H \,,
 \label{non-commuting}
 \\
\tilde H= -  \tau_1\epsilon(\k)  + \tau_3 [\sigma_x f_x(\k) +
\sigma_yf_y(\k) ] \,.
 \label{commuting}
\end{eqnarray}

The ``Hamiltonian'' $\tilde H$ in Eq. (\ref{commuting})
anticommutes both with $\cP=\tau_2$ ($P^2=1$) and
$\cC=\tau_3\sigma_z$ ($C^2=1$), while the whole Green function
(\ref{non-commuting}) anticommutes with
$fk_z\tau_2+\omega\tau_3\sigma_z$. As a result there exist 3D
slices $(\omega=t \sin\alpha, fk_z=t\cos\alpha)$ with fixed
$\alpha$ (shown in Fig.\ref{Fig:Planes}c), where $\tilde H$
anticommutes with the constant matrix
\begin{equation}
\cQ_\alpha=  \tau_2 \cos\alpha + \tau_3\sigma_z
\sin\alpha\,,~~\cQ_\alpha^2=1\,.
 \label{tilde-tau}
\end{equation}
The topological charge for a given parameter $\alpha$ is
\begin{equation}
 N_{\alpha}= \frac{e_{ijl}}{24\pi^2} ~
\mathop{\bf tr} \int    d^2k ~dt ~\cQ_\alpha \tilde
G\partial_{k_i} \tilde G^{-1} \tilde G\partial_{k_j} \tilde
G^{-1}\tilde G\partial_{k_l}  \tilde G^{-1}\,,
\label{TopInvariant_alpha}
\end{equation}
where $k_i=(t,k_x,k_y)$ and
 \begin{eqnarray} \nonumber
 \tilde G^{-1}(t,k_x,k_y|\alpha)= t(\tau_2\sin\alpha - \tau_3\sigma_z  \cos\alpha)
 +\\
 \tau_1\epsilon (\k) - \tau_3 [\sigma_x f_x (\k) +
 \sigma_yf_y (\k)]\,,
 \label{tilde-G}
 \end{eqnarray}
 where $\k=(k_x,k_y,f^{-1}t\cos\alpha)$. When the parameter $\alpha$ changes from 0 to $\pi/2$, the
topological charge (\ref{TopInvariant_alpha}) transforms from
Eq.(\ref{TopInvariant_B}) for the $(3+1)$-dimensional $^3$He-B to
Eq.(\ref{TopInvariant_P}) for the $(2+1)$-dimensional planar
phase. Naturally, along this path the topological invariant is
$N_\alpha=2n$ including the cases $\alpha=0,\pi/2$, so that
$N_{\rm B}=N_{\rm P}$.

The constructed connection between topological properties of
$^3$He-B and the planar phase can be generalized to include all
Hamiltonians within the DIII symmetry class as we show in the next section.

\subsection{Bulk-boundary correspondence for general DIII topological superconductors}

 In Sec.\ref{sec:bbint}A we have discussed the index theorem for a
subclass of DIII topological superconductors in 3D
described by Hamiltonians, similar to that of $^3$He-B. In
particular, we assumed that the Hamiltonian at $k_z=0$
cross-section of momentum space is equivalent to that of the
planar phase, which has a topological invariant protected by an
additional symmetry. However, this is not the case for the general
 Hamiltonian of class DIII. The reduction to $k_z=0$
shown in Fig.\ref{Fig:Reduction}b and in general to any time
reversal invariant plane (\ref{Eq:Line}) produces in this case a
DIII topological superconductor in 2D. As discussed in
Sec.~\ref{sec:Bphase}, the only symmetries that exist in general
for the 2D case are the TRS and PHS, which allow only the $\Z_2$
classification given by $\nu=(N_{\rm B}/2) \mod 2$.

The proof can be constructed as follows. First, we note that the
set of Hamiltonians (\ref{Eq:HeBHamiltonian-gen},
\ref{Eq:HigherInvariants}) with $n\in \Z$ contains representatives
from all topological classes of DIII symmetric Hamiltonians. Thus
we can continuously transform any given DIII Hamiltonian to one of
this set, preserving the value of $N_{\rm B}$. This generates a
deformation of the 2D Hamiltonian in the $k_z=0$ cross-section to
that of the generalized planar phase
(\ref{Eq:planarHamiltonian-gen},\ref{Eq:HigherInvariants}). This
deformation does not change the value of the $\Z_2$ invariant,
which coincides with that of the
 generalized planar state: $\nu=(N_{\rm B}/2) \mod 2= (N_{\rm P}/2 )\mod 2$
 in accordance with the dimensional reduction arguments of
 Sec.\ref{sec:bbint}B.

Finally, we note that the choice of the cross section $k_z=0$ is
arbitrary. Instead, we can choose any time-reversal invariant
plane of the form (\ref{Eq:Line}),  which allows one to investigate the
properties of bound states at the $y=0$ interface between a topological
superconductor of class DIII and a non-topological insulator. In
this case all the 2D Hamiltonians describing quasiparticles in
each cross section (\ref{Eq:Line}) have the same $\Z_2$ invariant
$\nu=(N_{\rm B}/2)\mod 2$. According to the bulk-boundary
correspondence in 2D time-reversal invariant topological
insulators a non-zero value of the $\Z_2$ invariant protects
topologically stable Kramers pairs of zero-energy surface
states\cite{KramersPairs}.
 Therefore we conclude that the bulk-boundary correspondence in DIII topological
 superconductors can be formulated as follows: Provided the value of the bulk 3D
invariant $N_{\rm B}/2$ is odd, the spectrum
$\varepsilon_b=\varepsilon_b(k_x,k_z)$ of surface states at the
boundary plane $y=0$ with a non-topological insulator has at least
one Kramers pair of topologically stable zero modes along each
line (\ref{Eq:Line}).

\section{Conclusion}
\label{sec:conclusion}

It is known that there exists a dimensional reduction, which connects
the classification of fully gapped  topological materials and the classification of nodal
systems with topologically protected zeroes in the energy spectrum (Fermi surfaces) described by
Ho\v{r}ava using the $K$-theory~\cite{Horava2005}.
An example of this connection is provided by the relation between the topological invariant
(\ref{TopInvariant_P}),
 which describes  the fully gapped 2+1 planar phase, and
the topological invariant that protects the point nodes in the
gapless 3+1 planar phase \cite{VolovikHeDroplet}. The invariants
are given by the same integral, but instead of integration over
the whole $(\omega,k_x,k_y)$-space in Eq.~(\ref{TopInvariant_P}), the
integral is taken over the sphere $S^3$ around the node in the
$(\omega,k_x,k_y,k_y)$-space.

Here we explicitly demonstrated a dimensional reduction, which
connects the fully gapped time reversal invariant topological
materials of different classes: the 2+1 planar phase of superfluid
$^3$He and the 3+1 superfluid $^3$He-B. As a result, two 3+1
topological systems become connected: the 3+1 gapless planar phase
and the gapful $^3$He-B. It is noteworthy that the planar phase of
$^3$He is topologically equivalent to the vacuum of the Standard
Model of particle physics in its massless phase of topological
semimetal \cite{NielsenNinomiya1981,VolovikHeDroplet,Creutz2008},
while the superfluid $^3$He-B is topologically equivalent to the
Standard Model vacuum in its massive phase of topological
insulator \cite{Volovik2010,Kaplan2012}. That is why the discussed
connection between the 3+1 topological states can be useful for
investigation of the topology of the Standard Model, which is also
supported by symmetry. The phenomenon discussed here, when the
discrete symmetry of the system leads to the continuous symmetry
of the single-particle Green function and correspondingly to an
integer topological invariant, is applicable to the vacuum of the
Standard Model.

\section{Acknowledgements} \label{Sec:Acknowledgements}
YM is grateful to J.E.~Moore and A.~Kitaev for valuable discussions.
MS and GEV acknowledge financial support  by the Academy of
Finland through its LTQ CoE grant (project $\#$250280).


\end{document}